\documentclass{article}
\begin{document}
\title{Embedding of the Brane into Six Dimensions}
\author{{\bf Merab Gogberashvili}\\
 Andronikashvili Institute of Physics, 6 Tamarashvili Str.,
Tbilisi 380077, Georgia \\
{\sl E-mail: gogber@hotmail.com }}
\maketitle
\begin{abstract}
Embedding of the brane metric into Euclidean (2+4)-space is found. Brane 
geometry can be visualized as the surface of the hyper-sphere in six 
dimensions which 'radius' is governed by the cosmological constant. 
Minkowski space in this picture is lied on the intersection of this surface 
with the plane formed by the extra space-like and time-like coordinates.

\end{abstract}
%%%%%%%%%%%%%%%%%%%%%%%%%%%%%%%%%%%%%%%%%%%%%%%%%%%%%%%%%%%%%%%%%%
\vskip 1cm

There has been great interest in recent years in models with extended extra 
dimensions. Ordinary gravity can be recovered if the observable universe is 
represented by a brane embedded in a higher-dimensional space with a 
non-factorizable geometry \cite{brane}. 

The useful method to study brane models can be the embedding theory 
\cite{eisenhart}. It is well known that $n$-dimensional space-time can be 
embedded into $N$-dimensional pseudo-Euclidean space with 
$n \le N \le n(n+1)/2$ \cite{eisenhart,friedman}. Thus, no more than ten 
dimensions are required to embed any 4-dimensional solution of Einstein's 
equations with arbitrary energy-momentum tensor. There also exists Campbell's 
theorem \cite{campbell}, which implies that any solution of $n$-dimensional 
Einstein's equations can be embedded, at least locally, in a space-time that 
is itself a solution of $(n+1)$-dimensional, vacuum Einstein's equations 
\cite{RTZ}. Several authors tried to interpreted the embedding as producing 
an effective stress-energy tensor in low dimensions \cite{PWM}.

The embedding procedure is also interesting from a purely mathematical point 
of view. It allows invariant classification of known solutions of Einstein's 
equations to be made \cite{exact}. Furthermore, the embedding method may lead 
to new solutions. For example, the maximal analytic extension of Schwarzschild
solution was independently found in this way \cite{fronsdal}. 

Embedding of the space-time with the coordinates $x^\alpha$ and metric
$g_{\alpha\nu}$ into pseudo-Euclidean space with the coordinates $X^A$ and 
with the flat metric $\eta_{AB}$ is given by
\begin{equation}\label{metric} 
ds^{2}= g_{\alpha\nu}dx^\alpha dx^\nu =
\eta_{AB}\partial_\alpha X^A \partial_\nu X^B dx^\alpha dx^\nu =
\eta_{AB}dX^AdX^B. 
\end{equation} 
For example, it is well known \cite{ros}, that Schwarzschild metric
\begin{equation} \label{sch} 
ds^{2}=\left(1-\frac{2m}{r}\right)dt^{2} - \frac{dr^{2}}{(1-2m/r)} - 
r^{2}d\Omega^{2} 
\end{equation} 
admits isometric embedding of class 2 into Euclidean (2+4)-space.
Embedding functions in this case are: 
\begin{eqnarray} \label{schEm}
X^{1}=\sqrt{1-\frac{2m}{r}}\cos t,~~~ X^{2}=\sqrt{1-\frac{2m}{r}}\sin t, 
~~~ X^{3}=f(r), \\ 
X^{4}=r\sin\theta\cos\varphi, ~~~~~~~
X^{5}=r\sin\theta\sin\varphi, ~~~~~~~ X^{6}=r\cos\theta, \nonumber
\end{eqnarray} 
where $f(r)$ is solution of the equation 
\begin{equation} \label{f}
f'^2 = \left(m^2 /r^4 + 2m/r\right)/\left(1 - 2m/r\right) . 
\end{equation}

In this paper we want to present embedding of the brane metric (which was 
introduced in \cite{brane}) into 6-dimensional pseudo-Euclidean space with 
the same signature $(2+4)$ as for the Schwarzschild case. Necessity of two 
time directions for embedding of $P$- and $M$-branes was shown in 
\cite{ADGHSP}.

We looking for the functions $X^A$ which fulfill the relation 
\begin{equation} \label{brane} 
ds^{2}= e^{2a\xi}dl^2 - d\xi^2 = dX_0^2 -
dX_1^2 - dX_2^2 - dX_3^2 + dX_\tau^2 - dX_\kappa^2. 
\end{equation} 
Here 
\begin{equation} \label{l}
l = \sqrt{t^2 - x^2 - y^2 - z^2}
\end{equation}
is the length in 4-dimensional Minkowski space-time, $\xi$ is the fifth 
coordinate orthogonal to the brane and $a$ is the parameter connected with 
the 5-dimensional cosmological constant $\pm \Lambda$. For the simplicity on 
the brane we considering Minkowski metric. Ricci tensor of the 5-dimensional 
space-time where the cosmological constant $\Lambda$ appears is not zero 
\cite{brane}, while we assume that bulk (2+4)-space to be pseudo-Euclidean 
again. This can be interpreted as a kind of geometrical introduction of the 
cosmological constant.

It can be checked that embedding (\ref{brane}) is done by the functions
\begin{eqnarray} \label{braneEm} 
X_\alpha = e^{a\xi} x_\alpha, \nonumber \\
X_\tau = \left( l^2 - \frac{1}{4}\right) e^{a\xi} - 
\frac{1}{a^2} e^{-a\xi}, \\
X_\kappa = \left( l^2 + \frac{1}{4}\right) e^{a\xi} - \frac{1}{a^2}
e^{-a\xi}, \nonumber
\end{eqnarray} 
where $x^\alpha$ are coordinates of Minkowski space-time and index $\alpha $ 
runs over $0, 1, 2, 3$. 

Inverse expression of 5-dimensional coordinates by the embedding functions
has the form 
\begin{eqnarray} \label{inverse} 
x_\alpha = \frac{X_\alpha}{2(X_\kappa - X_\tau)}, \nonumber \\
\xi = a \ln [2(X_\kappa - X_\tau)] .
\end{eqnarray} 

The geometry of 5-dimensional metric (\ref{brane}) can be visualized as
the surface of the hyper-sphere in six dimensions, since 
\begin{equation} \label{sphere} 
X_0^2 - X_1^2 - X_2^2 - X_3^2 +
X_\tau^2 - X_\kappa^2 = \frac{1}{a^2}. 
\end{equation} 
The radius of this 'sphere' is $1/|a|$ and thus governed by the value of the 
5-dimensional cosmological constant $\pm \Lambda $. In this picture 
4-dimensional Minkowski space-time is the intersection of this hyper-sphere 
with the plane 
\begin{equation} \label{plane} 
X_\kappa - X_\tau = \frac{1}{2} , 
\end{equation} 
where $X_\kappa$ and $X_\tau$ are respectively extra space-like and time-like 
coordinates of the 6-dimensional space-time. 

Possibly here we have correlations with the situation of the linearization 
of conformal group, the symmetry group of the main equations of physics in 
zero-mass limit. A long time ago it was discovered that the non-linear 
15-parameter conformal transformations can be written as a linear 
Lorentz-type transformation in a (2+4)-space. For this case the intersection 
of the null 6-cone with the null 5-plane forming by the extra time-like and 
space-like coordinates has induced metric of the Minkowski form (for these 
subjects see, for example, \cite{PR}) and formulae of this embedding are 
similar to (\ref{inverse}) - (\ref{plane}) we have in the branes case. 

At the end of the paper we want to note that (2+4)-space is the interest 
object for Kaluza-Klein models. For compact extra dimensions this space was 
studied in \cite{PI} and in the context of brane models with non-factorizable 
geometry in \cite{GM}.

\vskip 0.5cm

{\bf Acknowledgements:} Author would like to acknowledge the hospitality
extended during his visits at the Abdus Salam International Centre for
Theoretical Physics where this work was done.

\end{document}